# Vulnerability Analysis of Azure Blockchain Workbench Key Management System

Dmitry Tanana
*Laboratory of Combinatorial Algebra*
*Ural Federal University*
Yekaterinburg, Russia
ddtanana@urfu.ru

*Abstract*—With rise of blockchain popularity, more and more people seek to implement blockchain technology into their projects. Most common way is to take existing blockchain stack, such as Azure Blockchain Workbench or Oracle Blockchain Platform. While the blockchain technology is well-protected by its algorithms it is still vulnerable because its privacy relies on regular cryptography. And mistakes or vulnerabilities in key management protocols can affect even the most secure blockchain projects. This article considers question of vulnerabilities within Azure Blockchain Workbench key management system. We describe potential threats for each stage of key management lifecycle based on public reports and then assess how likely are those threats to realize within Azure Blockchain Workbench environment based on the technical documentation for Azure Blockchain Workbench and Azure Key Vault. Finally, we compile results of our assessment into the key management threat table with three distinct degrees of protection – fully protected, partially protected and not protected.

I. INTRODUCTION

Nowadays one of the most discussed topics in the field of information technology are distributed ledger systems, also known as blockchain. Such systems are being introduced around the world in many areas including medicine, education, finance and many others.

End user security in those systems relies primarily on cryptographic encryption, therefore the blockchain is only as secure as the encryption keys used. Indeed, many well-known incidents in which data from blockchain was somehow disclosed, happened only because of key mismanagement [1]. Therefore, the safety of information in distributed ledger systems directly depends on the key strength and the effectiveness of mechanisms and protocols associated with their management. Our paper will focus on the vulnerability analysis of blockchain key management systems on the example of Azure Blockchain Workbench. To the best of our knowledge, this would be the first analysis of this type for Azure Blockchain Workbench.

II. CRYPTOGRAPHIC KEY LIFECYCLE STAGES AND THREATS

*A. Lifecycle*

In order to evaluate specific security measures, we must consider the whole lifecycle of cryptographic key within the distributed ledger system.

The cryptographic key lifecycle is a set of states through which the key passes during its existence in automated system.

Cryptographic key lifecycle can be divided into following stages [2], as seen on Fig. 1:

• Generation – key is generated with required cryptographic properties.

• Registration – key is registered in the database and issuing key certificate.

• Distribution – keys are distributed to the owners via secure channel.

• User initialization – user is authenticated for the purpose of key reception.

• Operation – keys are used to protect the information during regular operations.

• Update – key is replaced with a new one after a certain period of time.

• Recovery – key is restored after being lost, if it was lost without compromise possibility.

• Cancellation – key is revoked as a result of expiration or compromise.

• Storage and archiving – on this stage keys are stored in proper conditions, ensuring their safety until replacement, discontinued keys are archived and stored until their complete removal.

• Erasure – key is completely deleted from the system after expiration or compromise.

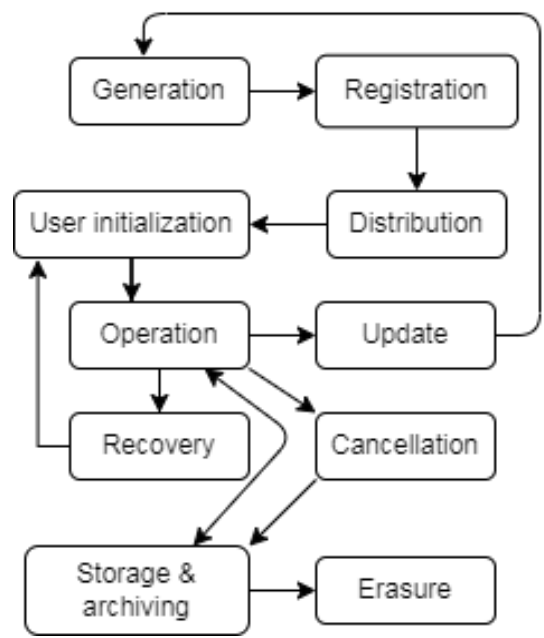

Fig. 1. Block diagram of a blockchain cryptographic key lifecycle stages.

This work was supported by the Ministry of Science and Higher Education of the Russian Federation, project no. FEUZ-2020-0016 .

*B. Threats*

We'll consider potential threats for each of the stages based on public reports about implemented attacks or attempted attacks.

We propose the following structure for describing threats:

• Name of the threat.

• Description of the threat.

• Mitigation measures to reduce the likelihood of the threat.

The main threats at the generation stage:

– Threat of generating weak cryptographic key [3]. This threat is a result of unsuccessful random number generator choice. It can produce a number which will result in generation of weak cryptographic key.

To mitigate this threat, strength of cryptographic keys must match the value of data it protects, and the period of time during which the data is protected by that key. Cryptographic key must be long enough for its purpose and generated using high quality random number generator, ideally the one that collects entropy from a suitable hardware source.

The main threats at the distribution and operation stages:

– Threat of insecure distribution of cryptographic keys [4]. Key distribution is one of the most vulnerable stages in the whole lifecycle. If this process is not secure enough, keys can easily become compromised.

To mitigate this threat only secure protocols should be used for the distribution of cryptographic keys and certain rules of conduct should be followed by end users (do not transfer keys by open channels, etc.).

– Threat of key misuse [4]. If cryptographic keys are used for tasks that do not correspond with its generation purpose, that may lead to problems with maintenance of required protection level protection.

To mitigate this threat each cryptographic key should be generated for one specific purpose only.

– Threat of key reuse [4]. Key being reused after its expiration can lead to compromise.

To mitigate this threat reuse of cryptographic keys should not be allowed. For a new operation new cryptographic key should be generated instead.

– Threat of using a single cryptographic key to encrypt large amounts of data [4]. Using one single cryptographic key to encrypt large amounts of data (at the same time or at multiple instances) can make this key more vulnerable for attacks.

To mitigate this threat a single cryptographic key should not be used for high amount of data encryption. If necessary, a new cryptographic key should be generated instead.

– Internal threats (attacker can have access to cryptographic keys) [5]. If attacker has constant access to the cryptographic keys, that access can be used for malicious purposes or keys can be transferred and/or sold to a third party.

To mitigate this threat, it is necessary to employ user authentication, dual control and separation of roles.

The main threats at the storage stage:

– Threat of improper storage [6]. Any bypass attack with filtration of protected data can compromise cryptographic keys.

To mitigate this threat cryptographic keys must never be stored along with data they are protecting (for example, on a same server, in a same database, etc.).

– Threat of weak protection [6]. The cryptographic keys, even if they are stored only in server memory, may be vulnerable to compromise.

To mitigate this threat: regardless of the storage, keys must be stored only in encrypted form (exception: if they are stored in an environment which is protected from unauthorized access).

– Threat of inaccessibility [6]. If cryptographic keys are not available when they are required or are lost due to any malfunction, and the backup is not available, data encrypted with these keys may also be inaccessible or lost.

To mitigate this threat storage location of cryptographic keys must be reliably protected from failure and backups should be kept for the keys themselves if they need to be recovered.

– Threat of logging absence [6]. If cryptographic key lifecycle is not fully recorded or logged it will be more difficult to determine when exactly the security incident happened, which will affect the provision of security information for other cryptographic keys in the future.

To mitigate this threat, it is necessary to keep full action logs for cryptographic keys. The journal should be protected from third-party access and ideally stored on a distributed ledger so it would be impossible to remove entries from it.

The main threats at the key erasure stage:

– Threat of leaving cryptographic keys intact after their expiration [7]. Keys which are not properly destroyed can be compromised and used by attackers.

To mitigate this threat cryptographic keys must be destroyed (safely removed without leaving any traces) after their expiration, if they are not clearly required for later use (for example, to decrypt data).

– Threat of manual management [7]. Manual management of cryptographic keys using paper or unsuitable electronic tools, such as Excel spreadsheets, and accompanying manual input of cryptographic keys will lead to mistakes that can make keys extremely vulnerable.

To mitigate this threat management of cryptographic keys should be entrusted to an automated system, which is designated for key management and corresponds to modern information security requirements.

Overall block diagram of cryptographic key management system threats with corresponding stages can be seen on Fig. 2.

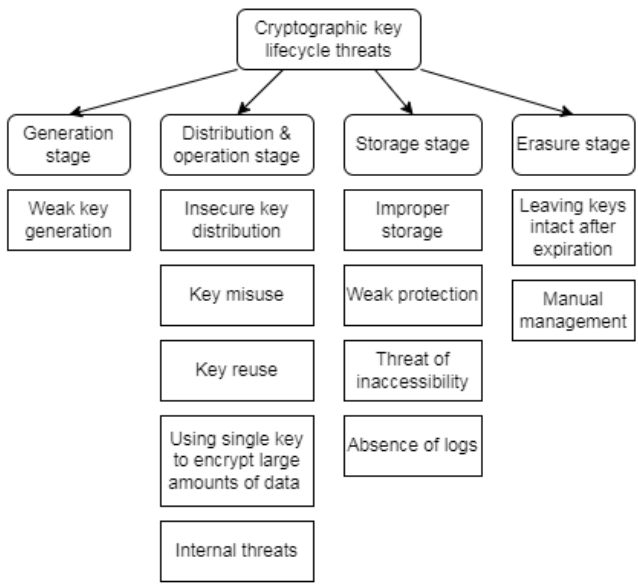

Fig. 2. Block diagram of a blockchain cryptographic key lifecycle threats with corresponding stages.

After analyzing potential threats to cryptographic keys management process, we will consider them specifically for the Azure Blockchain Workbench [8].

### III. AZURE BLOCKCHAIN WORKBENCH SECURITY SPECIFICATIONS

The Azure Blockchain Workbench key management system is based on the Azure Key Vault [9]. Blockchain participants are using Azure Key Vault or third-party certificate authority in order to create identifiers and associated certificates which are required for all nodes to carry out transactions in blockchain network.

Here are the rules and recommendations described in Azure Blockchain Workbench documentation [8,9] for cryptographic key management.

Generation Stage:

• Generation of public and private keys occurs according to the EC-HSM (Elliptic Curve - Hardware Security Module) algorithm with P-256 elliptic curve key in base version.

• If the private key is lost and cannot be restored or updated, then it will be necessary to generate a new private key and register a new identity at the certificate authority.

Storage stage:

• Keys are stored in the browser cache when they are initially added to the participant's blockchain wallet so that the console could use them to control blockchain components.

• Customers are encouraged to export keys and import them into their own cryptographic key management system in case they clear browser cache or switch browsers.

• Customers are responsible for storage, backup and recovery of all exported keys.

For generating and storing a cryptographic key Azure Blockchain Workbench also offers the use of HSM (Hardware Security Module). It complies with the FIPS 140-2 Level 3 standard. The blockchain network node is responsible for the correct configuration and operation of the HSM.

Also in the HSM case the following cryptographic key management rules are used [6]:

• The private key for the node should not be stored in browser cache.

• The private key must be accessible from the HSM via proxy.

• When registering administrator IDs of another node or client application in certification authority using the console, their private keys should not be stored in the HSM because they need the private key to operate in the network.

### IV. RESULTS AND CONCLUSION

Based on the analysis of Azure Blockchain Workbench, technical documentation for Azure Blockchain Workbench and Azure Key Vault and the compiled description we will analyze threats for Azure Blockchain Workbench key management system and assess the degree of protection from them.

In our study we'll use the following degrees of protection:

• Fully protected – security measures of Azure Blockchain Workbench completely eliminate the possibility of threat.

• Partially protected – security measures of Azure Blockchain Workbench partially (or completely, but not in all cases) eliminate the possibility of a threat. The likelihood of a threat is still present.

• Not protected – security measures of Azure Blockchain Workbench for this threat do not exist or do not fulfill their functions. High likelihood of a threat present.

The results of our threat analysis can be seen in table 1.

TABLE I. KEY MANAGEMENT THREAT TABLE FOR AZURE BLOCKCHAIN WORKBENCH

| № | Threat | Security element of Azure Blockchain key management system for this threat | Degree of protection |
|---|---|---|---|
| 1 | Threat of generating weak cryptographic key | Generation of public and private keys occurs according to EC-HSM algorithm with P-256 elliptic curve key in base version | Fully protected |
| 2 | Threat of key misuse | Not found | Not protected |
| 3 | Threat of key reuse | Key management rules | Fully protected |
| 4 | Threat of using a single cryptographic key to encrypt large amounts of data | Not found | Not protected |
| 5 | Threat of improper storage | Key management system or certified Hardware Security Module | Fully protected |
| 6 | Threat of weak protection | Key management system or certified Hardware Security Module | Fully protected |
| 7 | Threat of insecure distribution of cryptographic keys | Key management system or certified Hardware Security Module | Fully protected |

| 8 | Threat of leaving cryptographic keys intact after their expiration | Not found | Not protected |
|---|---|---|---|
| 9 | Internal threats (attacker can have access to cryptographic keys) | Certified Hardware Security Module system. No protection if Hardware Security Module is not used | Partially protected |
| 10 | Threat of inaccessibility | Clients are responsible for storage, backup and emergency recovery of all exported keys | Partially protected |
| 11 | Threat of logging absence | Certified Hardware Security Module system. No protection if Hardware Security Module is not used | Partially protected |
| 12 | Threat of manual management | Key management system or certified Hardware Security Module | Fully protected |

As a result of the threat analysis for the Azure Blockchain Workbench, we were able to assess the degree of protection from various attack vectors within its key management system and identify potential threats for the information security of Azure Blockchain Workbench key management system.

We'd like to underline that for many attack vectors protection relies solely on certified Hardware Security Module, which might be unavailable for broader user base. Additionally, for a few potential vulnerabilities – using single cryptographic key to encrypt large amounts of data and leaving cryptographic key intact after its expiration there were no security measures at all, which means that end users must be extra careful to prevent those.

Identified threats can be used by Azure Blockchain developers to further mitigate them or by end users in their risk minimization strategy.

REFERENCES

[1] P. Gaudry, A. Golovnev, "Breaking the encryption scheme of the Moscow Internet voting system", Available at: https://arxiv.org/abs/1908.05127.
[2] S. Rafaeli, D. Hutchison, 2003, "A Survey of Key Management for Secure Group Communication", ACM Computing Surveys, vol. 35, no. 3, pp. 309-329.
[3] F. Casino, T.K. Dasaklis, C. Patsakis,. 2019, "A systematic literature review of blockchain-based applications: Current status, classification and open issues", Telematics and Informatics, vol. 36, pp. 55-81.
[4] M. Ma, G. Shi, F. Li, 2019, "Privacy-Oriented Blockchain-Based Distributed Key Management Architecture for Hierarchical Access Control in the IoT Scenario", IEEE Access, vol. 7, pp. 34045-34059.
[5] L. Veltri, S. Cirani, S. Busanelli, G. Ferrari, 2013, "A novel batch-based group key management protocol applied to the Internet of Things", Ad Hoc Networks, vol. 11, no. 8, pp. 2724-2737.
[6] N.Z. Aitzhan, D. Svetinovic, 2018, "Security and Privacy in Decentralized Energy Trading Through Multi-Signatures, Blockchain and Anonymous Messaging Streams", IEEE Transactions on Dependable and Secure Computing, vol. 15, no. 5, pp. 840-852.
[7] Y. Sun, W. Trappe, K.J.R. Liu, 2004, "A scalable multicast key management scheme for heterogeneous wireless networks", IEEE/ACM Transactions on Networking, vol. 12, no. 4, pp. 653-666.
[8] "Azure Blockchain Workbench Architecture", Available at: https://docs.microsoft.com/en-us/azure/blockchain/workbench/architecture
[9] "About Azure Key Vault", Available at: https://docs.microsoft.com/en-us/azure/key-vault/general/overview